\documentclass[aps,amsmath,amssymb,prl,reprint,groupedaddress]{revtex4-2}

\usepackage{graphicx}
\usepackage{dcolumn}
\usepackage{bm}
\usepackage{hyperref}
\hypersetup{colorlinks=true,citecolor=blue,urlcolor=blue,linkcolor=blue}

\begin{document}


\title{A general framework of canonical quasinormal mode analysis for extreme nano-optics}

\author{Qiang Zhou}
\affiliation{School of Physics and Wuhan National Laboratory for Optoelectronics, Huazhong University of Science and Technology, Luoyu Road 1037, Wuhan 430074, People's Republic of China}

\author{Pu Zhang}
\email[Corresponding author: ]{puzhang0702@hust.edu.cn}
\affiliation{School of Physics and Wuhan National Laboratory for Optoelectronics, Huazhong University of Science and Technology, Luoyu Road 1037, Wuhan 430074, People's Republic of China}

\author{Xue-Wen Chen}
\email[Corresponding author: ]{xuewen\_chen@hust.edu.cn}
\affiliation{School of Physics and Wuhan National Laboratory for Optoelectronics, Huazhong University of Science and Technology, Luoyu Road 1037, Wuhan 430074, People's Republic of China}

\begin{abstract}
Optical phenomena associated with extremely localized field should be understood with considerations of nonlocal and quantum effects, which pose a hurdle to conceptualize the physics with a picture of eigenmodes. Here we first propose a generalized Lorentz model to describe general nonlocal media  under linear mean-field approximation and formulate source-free Maxwell's equations as a linear eigenvalue problem to define the quasinormal modes. Then we introduce an orthonormalization scheme for the modes and establish a canonical quasinormal mode framework for general nonlocal media. Explicit formalisms for metals described by quantum hydrodynamic model and polar dielectrics with nonlocal response are exemplified. The framework enables for the first time direct modal analysis of mode transition in the quantum tunneling regime and provides physical insights beyond usual far-field spectroscopic analysis. Applied to nonlocal polar dielectrics, the framework also unveils the important roles of longitudinal phonon polaritons in optical response. 
\end{abstract}

\maketitle

\emph{Introduction.}---
The research field of nano-optics has flourished along the development of concepts and techniques to shrink light to scales well below diffraction limit  \cite{Ebbesen2003SurfacePlasmons,Brongersma2010Plasmonics,Novotny2012NanoOptics,Polman2015Nanophotonics,Sandoghdar2020}. The subwavelength confinement primarily results from plasmonic oscillations in metal nanostructures \cite{Ebbesen2003SurfacePlasmons,Brongersma2010Plasmonics} or phononic oscillations in polar dielectric nanomaterials \cite{Keilmann2002,Glembocki2015}. Rapid advances in nanotechnologies have allowed sculpting structural morphology at nanometer and even subnanometer scales \cite{Norris2009,Cirac2012Probing,Dionne2012QuantumPlasmon,Hecht2012AtomicScale,Baumberg2015}, pushing field concentration towards extreme \cite{Baumberg2019ExtremeNanophotonics}. Recent experimental demonstrations suggest optical field could even be confined to cubic-nanometer volumes \cite{Zhang2013ChemicalMapping,Baumberg2016ScienceSingleMolecule,Apkarian2019vibrationalmodes,Dong2020photoluminescence}. Along this line of research, the field of nano-optics enters a new regime, where classical local treatment of nanomaterials becomes invalid. As an example, metallic gaps of a few nanometers or smaller exhibit significant nonclassical effects from electron nonlocality \cite{Cirac2012Probing,Stenger2015}, spill-over at metal surfaces \cite{Hongxing2015ResonanceShifts,Ding2017Plasmonic}, Landau damping \cite{cirac2017CurrentDependent,Khurgin2017LandauDamping} and quantum tunneling \cite{Savage2012,Dionne2013}. To describe these effects for plasmonic nanostructures ($\sim$100 nm) with tiny subfeatures by affordable computational resources, researchers have developed various effective models at different levels of approximation \cite{Hongxing2015ResonanceShifts,Ding2017Plasmonic,cirac2017CurrentDependent,Aizpurua2012,Yan2015,Crozier2016,Christensen2017,Yang2019NatureFramework}. For polar dielectric nanostructures, nonlocal responses also have been treated to properly describe the nanoscale physics \cite{Cardona1992,Arakawa1998,DeLiberato2020}.

While the system optical response could be computed by discretizing Maxwell’s equations in media with a suitable model, the interpretation of the physics and characterization of light-matter interaction properties are not straightforward. If the governing eigenmodes of the system were known, then system response could often be conceptualized. Thus the ability to obtain and orthonormalize the eigenmodes is essential to promote the development of extreme nano-optics. Both the openness and dissipative nature of the system call for a quasinormal mode (QNM) analysis, which has been established for classical local media \cite{Leung1994,Muljarov2011,Kristensen2012,Lalanne2013PRL,Yan2018PRB,Binkowski2018RieszProjection,Muljarov2016PRB,Lalanne2018LPR,Kristensen2020}. There have been few attempts to extend the QNM analysis to include electron nonlocality under a hydrodynamic treatment \cite{Stephen2017NonlocalQNM,Binkowski2019ModalAnalysis} and recently to incorporate a quantum hydrodynamic model (QHDM) to study extremely localized modes \cite{Li2021}. However, none of the extensions have attempted to construct an orthonormalization scheme for the QNMs, which is challenging since the aforementioned quantum effects have to be properly treated in the relation. Orthonormalization is crucial as it enables constructing system response through modal contributions for an arbitrary excitation and a direct evaluation of mode volume of each mode to facilitate quantum-optical studies \cite{Lalanne2018LPR,Muljarov2016PRB,Kristensen2020,Richter2019,Lalanne2021}.

Here we propose a generalized Lorentz model to describe general nonlocal media and formulate source-free Maxwell’s equations in the media as a linear eigenvalue problem (LEVP) to canonically define the QNMs. Then we introduce a general scheme to orthonormalize the eigenmodes and consequently establish a framework of canonical QNM analysis. Taking QHDM for metal and nonlocal polar medium for dielectric as examples, we present explicit QNM formalisms in corresponding media. We reveal the mode evolution process of a plasmonic dimer in the quantum tunneling regime where classical local treatment fails completely \cite{cirac2017CurrentDependent,Ding2018PRB} and show a mode transition occurs at a smaller gap than expected from extinction spectra. Moreover, we employ QNM analysis to interpret the optical responses of a silicon carbide (SiC) nanosphere influenced by longitudinal phonon modes. The examples cover a broad range of situations in extreme nano-optics. 

\emph{Generalized Lorentz model and LEVP formulation.}---
Classical local optical responses of materials can often be described by the Drude model or Lorentz oscillator model of charged particles with single or multiple resonances \cite{Taflove05}. Nonlocal responses of materials conspicuously manifest via longitudinal density waves of the constituent charged particles with a characteristic length much shorter than vacuum wavelength \cite{Hanke1978}. Inspired by the nature of nonlocality as interaction induced by charge density gradient, we propose a generalized Lorentz model (GLM) to incorporate various nonlocal responses under weak excitation
\begin{equation}\label{Eq:1}
\rho\ddot{\mathbf{X}}
+\widehat{\Gamma}_\mathrm{x}\dot{\mathbf{X}}
+\widehat{\Theta}_\mathrm{x}\mathbf{X}
= (q_\mathrm{p}/m_\mathrm{p})\rho\mathbf{E},
\end{equation}
where $\mathbf{X}$ is the relative displacement and the restoring and damping force constants have become operators containing spatial gradients. $\rho$ is the stationary charge number density normalized by the average number density ($n_\mathrm{u}$) and $q_\mathrm{p}/m_\mathrm{p}$ is the charge-mass ratio. A nonuniform $\rho$ covers situations of electron spill-over \cite{Crozier2016}, dynamic carrier control \cite{Brongersma2017} and gradient-alloyed semiconductors \cite{Nie2003}. The material response couples with Maxwell's equations through electric polarization $\mathbf{P}=n_\mathrm{u}\rho q_\mathrm{p}\mathbf{X}$. Assuming a time convention of $\exp(-i\widetilde{\omega}t)$ with $\widetilde{\omega}$ being a complex frequency, the equation for $\mathbf{P}$ in frequency domain reads
\begin{equation}\label{Eq:PJfreq}
i\rho\varepsilon_0\omega^2_\mathrm{p}\mathbf{E}-i\rho\widehat{\Theta}\mathbf{P}-i\rho\widehat{\Gamma}\mathbf{J}=\widetilde{\omega}\mathbf{J},
\end{equation}
with the polarization current $\mathbf{J}=-i\widetilde{\omega}\mathbf{P}$, modified restoring and damping force operators of $\widehat{\Theta}=\rho^{-1}\widehat{\Theta}_\mathrm{x}\rho^{-1}$ and $\widehat{\Gamma}=\rho^{-1}\widehat{\Gamma}_\mathrm{x}\rho^{-1}$. $\omega_\mathrm{p}=[n_\mathrm{u}q^2_{\mathrm{p}}/\!(\!m_\mathrm{p}\varepsilon_0\!)]^{1\!/\!2}$ resembles the plasma frequency. 
Equations (\ref{Eq:1}-\ref{Eq:PJfreq}) establish a Lorentz-type operator description of spatially dispersive media $\mathbf{E}=\widehat{\mathcal{L}}\mathbf{P}$ at optical frequencies, in parallel to the usual electric susceptibility descriptions of $\chi(\omega,\mathbf{r},\mathbf{r'})$ and $\chi(\omega,k)$ in real-space and wavevector-space forms respectively \cite{Landau84}. The GLM is basic since it is only based on the generic nature of material nonlocality and Newton’s second law. As listed in Table \ref{Table1} and detailed in SM \cite{SM}, our GLM accommodates at least the following nonlocal models, \textit{i.e.,} hard-wall hydrodynamic model \cite{Raza2011}, generalized nonlocal optical response \cite{Mortensen2014nonlocal}, QHDM \cite{Hongxing2015ResonanceShifts,cirac2017CurrentDependent,Ding2017Plasmonic} and nonlocal polar dielectrics \cite{DeLiberato2020}. The local response approximation is a special case.

\par With the materials described by the GLM, the source-free Maxwell's equations can be formulated as an LEVP
\begin{align}\label{Eq:3}
\mathcal{H}\Phi
\equiv\begin{bmatrix}
0&\frac{i}{\varepsilon_0\varepsilon_\infty}\nabla\times&0&-\frac{i}{\varepsilon_0\varepsilon_\infty}\\
-\frac{i}{\mu_0}\nabla\times&0&0&0\\
0&0&0&i\\
i\rho\varepsilon_0\omega_\mathrm{p}^2&0&-i\rho\widehat{\Theta}&-i\rho\widehat{\Gamma}
\end{bmatrix}\Phi
=\widetilde{\omega}\Phi
\end{align}
Here $\Phi=[\mathbf{E},\mathbf{H},\mathbf{P},\mathbf{J}]^\mathrm{T}$ and $\widetilde{\omega}$ are the eigenvector and eigenfrequency, respectively. $\varepsilon_0\varepsilon_\infty$ is the non-resonant background permittivity. In open space, the radiation boundary condition ($\mathbf{E}(\mathbf{r})\propto r^{-1}e^{i\widetilde{\omega}r/c}$ as $r\to\infty$) should be imposed. Equation \eqref{Eq:3} is formally similar to the auxiliary-field formulations developed for normal modes \cite{Fan2010PRL} and extended to treat losses for QNM analysis \cite{Yan2018PRB}. 
The original contribution here lies in the promotion of constants to operators $\widehat{\Theta}$, $\widehat{\Gamma}$ and the use of a nonuniform $\rho$ to describe general nonlocal responses and to develop a corresponding canonical QNM theory. 

\begin{table}[tb]
	\renewcommand{\arraystretch}{1.2}
	\centering
	\begin{tabular}{ lcccc }
		\hline
		\hline
		\rule{0pt}{11pt}
		Model & $\rho$ & $\widehat\Theta$ & $\widehat\Gamma$ & B.C. \\
		\hline
		LRA &1 &$\omega_0^2$ &$\gamma$ &/ \\
		HDM \cite{Raza2011} &1 &$-\beta^2\nabla(\nabla\cdot)$ &$\gamma$ &$\mathbf{n\cdot P}$ \\
		GNOR \cite{Mortensen2014nonlocal} &1 &$-(\beta^2\!+\!\gamma{D_{\!f}}\!)\nabla(\nabla\cdot)$\, &\,$\gamma\!-\!D_{\!f}\!\nabla(\nabla\cdot)$\, &$\mathbf{n\cdot P}$ \\
		QHDM &\,$\rho(\mathbf{r})\,$ &$-\widehat\Pi\!-\!\widehat\Sigma_2$ &$\gamma/\!\rho\!-\!\widehat\Sigma_1$ &/ \\
		Polar diel.\,\, &1 &Eq.\,\eqref{Eq:ThetaDielectric} &$\gamma_\mathrm{d}$ &\,\,$\mathbf{n}\cdot\overline{\tau}\mathbf{P}$ \\
		\hline
		\hline
	\end{tabular}
	\caption{List of nonlocal models formulated in GLM \cite{SM}.
		\label{Table1}}
\end{table}

\emph{General framework of canonical QNM analysis.}---
A canonical QNM theory should include a scheme to orthonormalize the modes. For normal modes of a Hermitian system, the normalization is carried out through an integration representing the field energy \cite{PhCv2,Fan2010PRL}. For QNMs, the exponential divergence of the mode field in the far field causes a difficulty for normalization. With continuous efforts \cite{Leung1994,Lalanne2013PRL,Kristensen2015,Muljarov2016PRB,Yan2018PRB,Lalanne2020}, the orthonormalization scheme for QNMs in classical local media has been established by resolving the divergence problem through a bilinear form and a complex coordinate mapping technique \cite{Lalanne2013PRL,Yan2018PRB}. However, the orthonormalization of QNM in nonlocal media remains an untouched challenging problem. By inspecting the orthonormalization formulas of Eq.\,(13) in Ref.\,\cite{Fan2010PRL} for normal modes and of Eq.\,(4) in Ref.\,\cite{Yan2018PRB} for QNMs, one sees each term in the expression for QNMs also has a close link to the mode energy. Energy-wise, for general nonlocal media, the difficulties are the proper treatments of various internal interaction energies of charged particles \cite{Hongxing2015ResonanceShifts,Ding2017Plasmonic,DeLiberato2020} and Landau damping as a type of interaction energy \cite{Mortensen2014nonlocal,cirac2017CurrentDependent}. Moreover, the nonuniform $\rho$ induces another barrier since this means a position-dependent plasma frequency and interaction energies. Our GLM formulation enables us to get around all these hurdles. Inspired by the close link to the field energy, we start with Poynting theorem \cite{Jackson1999} by evaluating the difference between the powers input by two current sources $I_\mathrm{c}=\int_{V}\!d\mathbf{r}\,[(i\mathbf{J}_{\mathrm{s},2})^*\cdot\mathbf{E}_1-\mathbf{E}^*_2\cdot i\mathbf{J}_{\mathrm{s},1}]$, where $\mathbf{E}_{i}$ is the response field of source $\mathbf{J}_{\mathrm{s},i}$ with \textit{i} = 1, 2. This leads to the expression for the electromagnetic energy in general nonlocal media \cite{SM}. Next we switch to the bilinear version (remove complex conjugate operations), e.g. $I_\mathrm{c}\to I_\mathrm{b}=\int_{V}\!d\mathbf{r}\,[i\mathbf{J}_{\mathrm{s},2}\cdot\mathbf{E}_1-\mathbf{E}_2\cdot i\mathbf{J}_{\mathrm{s},1}]$. By replacing the current sources with the fields, the bilinear form of Poynting theorem can be arranged to  \cite{SM}
\begin{align}\label{Eq:Poynting}
&
\left(\omega_1\!-\!\omega_2\right)\!\!\!
\int_V\!\!\!\!d\mathbf{r}\bigg(\!\!
\varepsilon_0\varepsilon_{\!\infty}\!\mathbf{E}_2\!\cdot\!\mathbf{E}_1
\!-\!\mu_0\mathbf{H}_2\!\cdot\!\mathbf{H}_1
\!+\!\frac{\mathbf{P}_{\!2}\!\cdot\!\widehat{\Theta}\mathbf{P}_{\!1}}{\varepsilon_0\omega_{\mathrm{p}}^2}
\!-\!\frac{\mathbf{J}_2\!\cdot\!\mathbf{J}_1}{\varepsilon_0\omega_{\mathrm{p}}^2\rho}
\!\!\bigg)
\notag\\
&-\!\!\frac{i}{\varepsilon_0\omega_{\mathrm{p}}^2}\!\!\int_V\!\!\!\!d\mathbf{r}\!
\left[\!\left(\!\mathbf{J}_2\!\cdot\!\widehat{\Gamma}\mathbf{J}_1
\!-\!\mathbf{J}_1\!\cdot\!\widehat{\Gamma}\mathbf{J}_2\!\right)
+
i\omega_1\!\!\left(\!\mathbf{P}_{\!2}\!\cdot\!\widehat{\Theta}\mathbf{P}_{\!1}
\!-\!\mathbf{P}_{\!1}\!\cdot\!\widehat{\Theta}\mathbf{P}_{\!2}\!\right)\!\right]
\notag\\
&-\!i\!\!\int_V\!\!\!\!d\mathbf{r}\,
\nabla\!\cdot\!\left(\mathbf{E}_1{\times}\mathbf{H}_2
-\mathbf{E}_2{\times}\mathbf{H}_1\right)
= I_\mathrm{b}.
\end{align}
Notice that the GLM formulation actually enables us to establish an unconjugated Lorentz reciprocity theorem \cite{PhCv2,Lalanne2013PRL} for general nonlocal media. 
We identify that $\widehat{\Theta}$ and $\widehat{\Gamma}$ should be real and symmetric under transposition from system energy considerations \cite{SM}, such as the positive definiteness of eigen-energies and the exponential decay caused by coupling to a continuum. The symmetry condition also implicates certain requirements on the boundary conditions for $\mathbf{P}$ across material interface \cite{Raza2011,DeLiberato2020,SM}. Such symmetry condition is indeed satisfied for all the situations of interest \cite{SM}. Consequently, the second integral of Eq.\,\eqref{Eq:Poynting} vanishes. The last integral of Eq.\,\eqref{Eq:Poynting} is normally converted to a surface integral at infinity and effectively brought to zero by using the perfectly matched layers \cite{Lalanne2013PRL}. However, a nontrivial surface integral may arise when unconventional electromagnetic boundary conditions are applied, e.g. in Feibelman's $d$-parameter model \cite{Feibelman1982,Yan2015,Christensen2017,Yang2019NatureFramework}. Although it could be treated by following the general spirit of our orthonormalization procedure \cite{Zhou2021QSR}, for the sake of clarity, here we restrict to volumetric media responses and thus drop the term. Then the left hand side of Eq.\,\eqref{Eq:Poynting} is left only with the first integral. Now considering the case that $\mathbf{E}_1$ and $\mathbf{E}_2$ are two sets of source-free ($\mathbf{J}_{\mathrm{s},1} = 0$ and $\mathbf{J}_{\mathrm{s},2} = 0$) eigenmode fields, the right hand side of Eq.\,\eqref{Eq:Poynting} becomes zero \cite{Lalanne2013PRL}. Thus  Eq.\,\eqref{Eq:Poynting} directly leads to the orthonormal relation $(\widetilde\omega_1\!-\!\widetilde\omega_2)(\!(\widetilde{\Phi}_2,\widetilde{\Phi}_1)\!)_{\mathcal{M}}=0$, where the bilinear form $(\!(\Psi_2,\Psi_1)\!)_{\mathcal{M}}\equiv\int{d}\mathbf{r}\,\Psi_2^\mathrm{T}\mathcal{M}\Psi_1$ is defined with
\begin{equation}\label{Eq:Moperator}
\mathcal{M}
= \mathrm{diag}\left\{\varepsilon_0\varepsilon_\infty,-\mu_0,(\varepsilon_0\omega_\mathrm{p}^2)^{\!-1}\widehat{\Theta},-(\varepsilon_0\omega_\mathrm{p}^2\rho)^{\!-1}\right\}.
\end{equation}
The normalization factors of QNMs in general nonlocal media are immediately obtained as $\mathcal{N}_m^2=(\!(\widetilde{\Phi}_m,\widetilde{\Phi}_m)\!)_\mathcal{M}$. $\mathcal{M}$ serves as a mapping operator such that the basis $\{\widetilde\Phi_m\}$ and its dual basis $\{\mathcal{M}\widetilde\Phi_n\}$ form a biorthogonal system. The completeness of the system is discussed in SM \cite{SM}. The GLM formulation and orthonormal relation constitute the core of the general framework of canonical QNM analysis and empower an analytical description of optical responses with numerically calculated QNMs. Analytical formulas can be derived by applying the orthonormal relation in parallel to the classical local theory \cite{Yan2018PRB,Lalanne2018LPR,SM}. Linear responses can be expanded as
$\Psi=\sum_m\alpha_m(\omega)\widetilde{\Phi}_m$
with coefficients
$\alpha_m(\omega)
= (\widetilde{\omega}_m-\omega)^{\!-1}
(\!(\widetilde{\Phi}_m,\mathcal{S})\!)_{\mathcal{M}}$, where $\mathcal{S}$ is the excitation source. The complex position-dependent mode volume reads 
$V_m=1/\{2\varepsilon_0n_\mathrm{d}^2\,[\widetilde{\mathbf{E}}_m(\mathbf{r}_\mathrm{d})\cdot\mathbf{u}_\mathrm{d}]^2\}$, which is evaluated for a dipolar emitter at $\mathbf{r}_\mathrm{d}$ along a unit vector $\mathbf{u}_\mathrm{d}$ in the medium with a refractive index of $n_\mathrm{d}$.

\emph{General framework applied to metals and dielectrics.}---
In the following, we take QHDM as the most sophisticated nonlocal model for metal and nonlocal polar medium for dielectric as two archetypal examples to explicitly work out their QNM theories. When metallic nanostructures are excited by light in the linear nonlocal response regime, the electron gas experiences additionally a pressure force arising from its internal energy \cite{Ding2017Plasmonic} and a viscoelastic force associated with Landau damping \cite{cirac2017CurrentDependent}. The two forces can be formulated as operators acting on the induced polarization $\mathbf{P}$ and current $\mathbf{J}$. Then the governing equation of QHDM can be written as \cite{SM}
\begin{equation}\label{Eq:operatorFormQHDM}
i\rho\varepsilon_0\omega_\mathrm{p}^2\mathbf{E}-i\rho(-\widehat{\Pi}-\widehat{\Sigma}_2)\mathbf{P}-i\rho(\gamma/\rho-\widehat{\Sigma}_1)\mathbf{J}=\widetilde{\omega}\mathbf{J}.
\end{equation}
$\gamma$ is the phenomenological damping rate. Comparison of Eq.\eqref{Eq:operatorFormQHDM} with Eq.\, \eqref{Eq:PJfreq} immediately reveals $\widehat{\Theta}=-\widehat{\Pi}-\widehat{\Sigma}_2$ and $\widehat{\Gamma}=\gamma/\rho-\widehat{\Sigma}_1$. The constituting operators are
\begin{subequations}
\begin{align}
\widehat{\Pi}
&= \nabla K_1(\nabla{\cdot})
- \nabla\nabla\cdot K_2\nabla(\nabla{\cdot}),
\\
(\widehat{\Sigma}_1)_{\bar{k}k}
&= \widehat{\overline{D}}_{\bar{j}}\left[
\eta'\,(\delta_{\bar{k}j}\delta_{\bar{j}k}+\delta_{\bar{k}k}\delta_{\bar{j}j})
+\xi'\,\delta_{\bar{k}\bar{j}}\delta_{kj}
\right]\widehat{D}_j,
\\
(\widehat{\Sigma}_2)_{\bar{k}k}
&= \widehat{\overline{D}}_{\bar{j}}\left[
\mu'\,(\delta_{\bar{k}j}\delta_{\bar{j}k}+\delta_{\bar{k}k}\delta_{\bar{j}j})
+\zeta'\,\delta_{\bar{k}\bar{j}}\delta_{kj}
\right]\widehat{D}_j,
\end{align}
\end{subequations}
where $\widehat{\mathbf{D}}=\nabla-\rho^{-1}\nabla\rho$, and $\widehat{\overline{\mathbf{D}}}=\nabla+\rho^{-1}\nabla\rho$.
$K_{1,2}$ and $\eta'$, $\mu'$, $\xi'$, $\zeta'$ are functions of $\rho$ \cite{SM}.
The operators comply with the requirement of being real symmetric. The canonical QNM theory for QHDM naturally follows from the general framework. A direct evaluation of $(\!(\widetilde{\Phi}_m,\widetilde{\Phi}_m)\!)_\mathcal{M}$ yields the normalization factor 
\begin{align}\label{Eq:NfactorQHDM}
\mathcal{N}_m^2
&=\!\! \mathcal\int\!d\mathbf{r}\,2\varepsilon_0\varepsilon_\infty\mathbf{\widetilde{E}}_m{\cdot}\mathbf{\widetilde{E}}_m
-\!\! \int\!\!d\mathbf{r}\,[i\gamma\widetilde{\omega}_m/(\varepsilon_0\omega_\mathrm{p}^2\rho)]\,\mathbf{\widetilde{P}}_m{\cdot}\mathbf{\widetilde{P}}_m
\notag\\
&- (1/\varepsilon_0\omega_\mathrm{p}^2)\!\!
\int\!\!d\mathbf{r}\,\mathbf{\widetilde{P}}_m{\cdot}(2\widehat{\Pi}+2\widehat{\Sigma}_2-i\widetilde{\omega}_m\widehat{\Sigma}_1)\mathbf{\widetilde{P}}_m.
\end{align}
In the expression, various nonlocal interactions are clearly manifested. The second integral accounts for the effect of nonuniform $\rho$. The second line encapsulates various nonlocal responses, such as the contributions of electron pressure ($\widehat{\Pi}$), elastic ($\widehat{\Sigma}_2$) and viscous ($\widehat{\Sigma}_1$) effects.

In polar dielectrics, nonlocality originates from ionic interactions through longitudinal phonons. By inspecting the dynamic equation for lattice vibrations proposed in Ref.\,\cite{DeLiberato2020}, it amounts to Eq.\,\eqref{Eq:PJfreq} with $\rho=1$, $\widehat\Gamma=\gamma_\mathrm{d}$ and
\begin{equation}\label{Eq:ThetaDielectric}
\widehat\Theta=\omega^2_\mathrm{T}
+\beta^2_\mathrm{L}\nabla(\nabla\cdot)
-\beta^2_\mathrm{T}\nabla{\times}(\nabla\times).
\end{equation}
$\omega_\mathrm{T}$ and $\beta_\mathrm{T}$ ($\beta_\mathrm{L}$) are the transverse optical phonon frequency and velocity of the transverse (longitudinal) phonons, respectively. The last two terms in Eq.\,\eqref{Eq:ThetaDielectric} follow from the divergence of the stress tensor $\overline{\tau}$, \textit{i.e.,} $\nabla\cdot\overline{\tau}=(\widehat\Theta-\omega^2_\mathrm{T})\mathbf{P}$. The nonlocal force operator is real symmetric with a required boundary condition $\mathbf{n}\cdot\overline{\tau}\mathbf{P}=0$ \cite{DeLiberato2020}. Therefore our general framework directly leads to a QNM theory for nonlocal polar dielectrics. The detailed derivations and numerical implementation of the two exemplary theories are provided in SM \cite{SM}.

\begin{figure}[tb]
\centering
\includegraphics[width=1.0\linewidth]{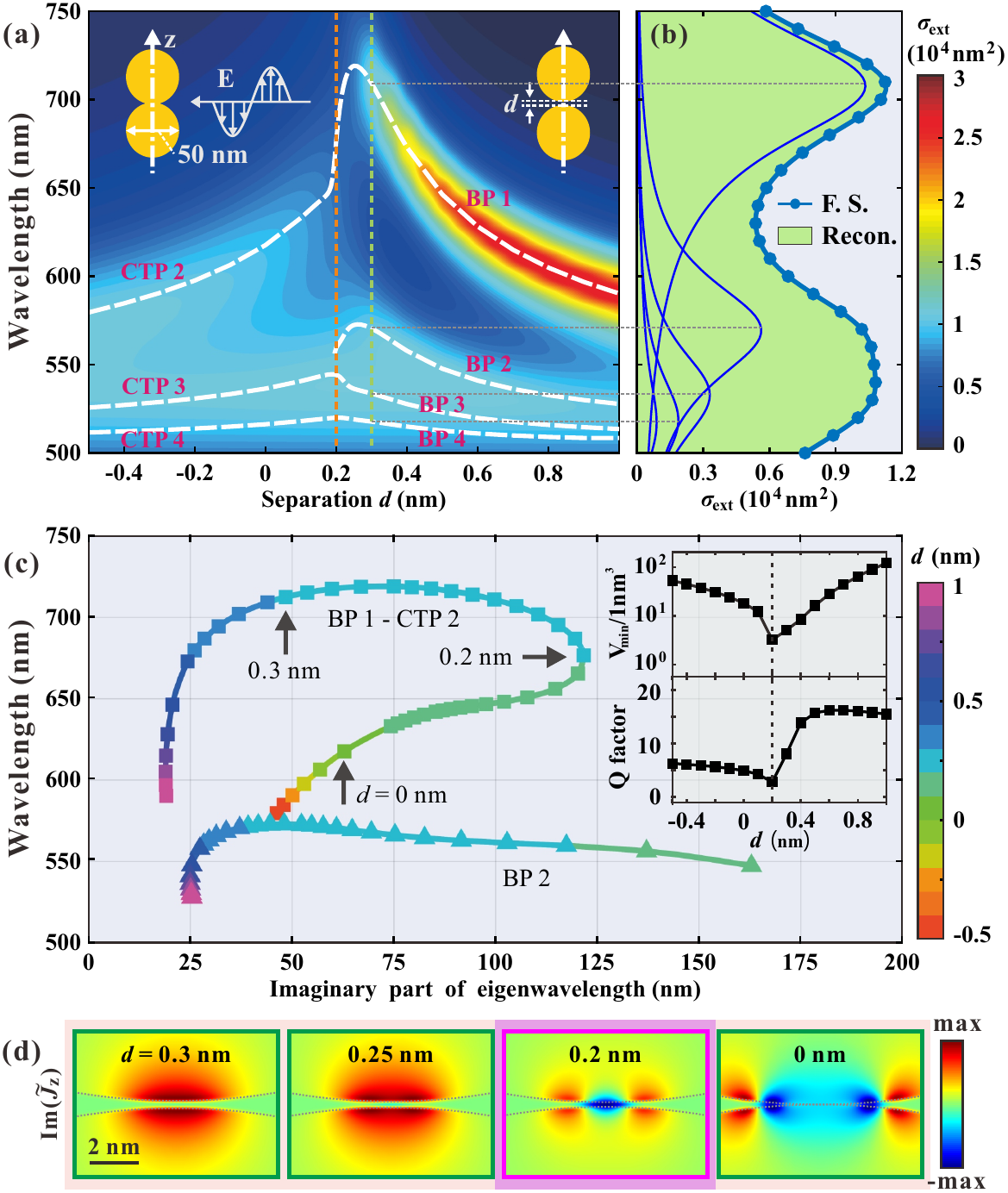}
\caption{(\textbf{a}) Extinction spectrum evolution with the gap size $d$ for a gold nanosphere dimer. The geometry and illumination scheme are illustrated in the inset. The white dashed traces indicate the real part of the resonant wavelengths obtained from QNM analysis. (\textbf{b}) QNM reconstruction of the extinction spectrum for $d=0.3$ nm. (\textbf{c}) Mode-evolution paths on the complex wavelength plane. The insets display minimum BP1/CTP2 mode volume and quality factor as functions of $d$. (\textbf{d}) The dominant component of BP1/CTP2 mode current density $\mathrm{Im}\{J_z\}$ at various gap sizes.}
\label{fig:fig1}
\end{figure}

\emph{Mode transition in quantum tunneling regime.}---
A metallic nanosphere dimer with an ultrathin gap is an excellent platform for studying intriguing quantum plasmonic phenomena such as quantum electron tunneling \cite{Savage2012,Dionne2013,Aizpurua2012,Borisov2012,Jensen2019,Li2016}. Considerable experimental and theoretical endeavors focus on the plasmon mode transition behavior as the gap gradually closes, but have not reached a complete consensus. A plausible reason is that the resonance information is indirectly retrieved from the far-field extinction or scattering spectra which strongly depend on the illumination and detection schemes. Our QNM theory provides an unprecedented opportunity to directly uncover the path of mode evolution in the quantum tunneling regime. Here we employ the QHDM-based QNM theory for the investigation since QHDMs with recent developments \cite{Hongxing2015ResonanceShifts,Ding2017Plasmonic,cirac2017CurrentDependent} can self-consistently describe various nonlocal and quantum effects for nanospheres down to 1 nm diameter \cite{cirac2017CurrentDependent} and to quantum tunneling regime where classical local treatment fails completely \cite{cirac2017CurrentDependent,Ding2018PRB}. As shown in Fig.\,\ref{fig:fig1}(a), we study a gold nanosphere (50 nm diameter) dimer and plot the extinction spectrum evolution map obtained from full numerical QHDM-based simulations \cite{SM}. The involved bonding plasmon (BP) and charge transfer plasmon (CTP) resonances are indicated. The first CTP is beyond 1 $\mu$m and not shown here. The map of extinction spectra in Fig.\,\ref{fig:fig1}(a) is in excellent qualitative agreement with the experimental spectra in Fig.\,2 of Ref.\,\cite{Savage2012}. As shown in more detail in SM \cite{SM}, good quantitative agreements are found by modeling the same geometry and illumination scheme as in Ref.\,\cite{Savage2012}. Then we perform QNM analysis to identify the eigenmodes of the system. As a benchmark for the orthonormal relation, we reconstruct the extinction spectrum for $d=0.3$ nm. Figure \ref{fig:fig1}(b) displays a perfect agreement between the full simulation and reconstruction with the contributions of 30 QNMs.

Previous far-field spectroscopic studies \cite{Savage2012} recognize the onset of quantum tunneling at the gap of 0.3 nm, which seems to be confirmed by our extinction spectra in Fig.\,\ref{fig:fig1}(a) (green dashed line). However, based on our QNM analysis, the mode evolution paths on the complex wavelength plane in Fig.\,\ref{fig:fig1}(c) indicate that the mode transition occurs at a smaller gap of 0.2 nm, coincident with the kink on the BP1-CTP2 dashed trace in Fig.\,\ref{fig:fig1}(a). It is better evidenced by the modal current profiles in Fig.\,\ref{fig:fig1}(d). On the threshold, current at the gap center emerges  and concomitantly the mode order changes. The mode has the smallest volume $V_\mathrm{min}$ as shown in an inset of Fig.\,\ref{fig:fig1}(c) \cite{SM}. Despite the accompanying lowest quality factor, we emphasize that  the optical responses under far-field excitation can still be conceptualized with the mode. The  extinction spectrum around the resonance (680 nm) are dominated by the BP1/CTP2 mode. The reconstruction of the spectrum in terms of modal contributions at $d=0.2$ nm, and the mode profile evolution for high-order modes are provided in SM \cite{SM}.

\begin{figure}[tb]
	\centering
	\includegraphics[width=1.0\linewidth]{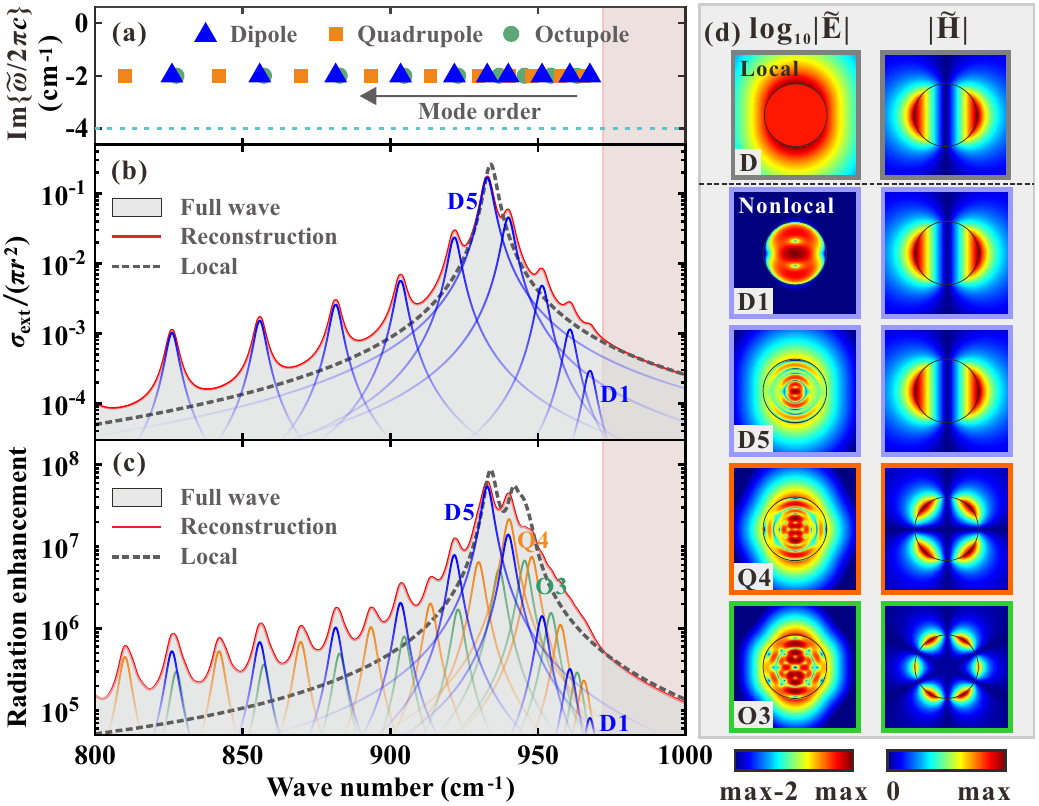}
	\caption{(\textbf{a}) QNM spectrum for a SiC nanosphere (diameter 10 nm; $\gamma_\mathrm{d} = 4$ cm$^{-1}$). (\textbf{b}) Extinction spectra of the nanosphere for a plane wave.  (\textbf{c}) Radiation enhancement spectra of an electric dipole placed 5 nm away from and perpendicular to the nanosphere surface. Results of local, nonlocal full-wave simulations and QNM reconstructions are plotted in dash-balck traces, gray shadings and red traces, respectively. The pinkish shading indicates the region without phonon polariton modes. (\textbf{d}) A gallery of modal electric and magnetic field profiles. See SM \cite{SM} for the details of calculations.}
	\label{fig:fig2}
\end{figure}

\emph{QNM analysis of a nonlocal polar dielectric nanoparticle.}---
The optical responses of polar dielectrics are classically characterized with the Fröhlich resonance. As the particle size shrinks to nanometer scale, nonlocal effects owing to longitudinal phonons become significant  \cite{Ratchford2019InfraredDielectricFunction,DeLiberato2020}. Here we perform QNM analysis for a SiC nanosphere (10 nm diameter) and unveil how the nonlocal responses are dictated by longitudinal phonon polaritons. Firstly the QNMs of the nanosphere are arranged according to the complex eigenfrequency as in Fig.\,\ref{fig:fig2}(a). Consistent with the dispersion of longitudinal phonons \cite{Karch1994SiliconCarbide}, the eigenfrequencies decrease with the mode order. Meanwhile the imaginary parts are essentially constant $\mathrm{Im}\{\widetilde\omega_m\}=-\gamma_\mathrm{d}/2$, which implies the electric fields of these QNMs are confined inside the nanosphere and largely longitudinal (left panel of Fig.\,\ref{fig:fig2}(d)). Their magnetic fields could be dipolar (D), quadrupolar (Q), octupolar (O) and etc (right panel of Fig.\,\ref{fig:fig2}(d)). 

Assuming a plane wave illumination, the extinction spectra are calculated for both nonlocal and local responses. As shown in Fig.\,\ref{fig:fig2}(b), the nonlocal corrections introduce extra resonances and can be perfectly reconstructed with modal contributions from 11 QNMs, confirming the validity of our orthonormal relation for nonlocal polar dielectrics. For the far-field plane wave excitation, only the dipolar modes are excited. By placing an electric dipole 5 nm away from the nanosphere, a wealth type of modes could be excited. Figure \ref{fig:fig2}(c) shows the radiation enhancement spectrum, which includes contributions from various dipolar, quadrupolar and octupolar modes. For clarity, we examine more closely the far-field response in Fig.\,\ref{fig:fig2}(b) to showcase the intriguing implications from our nonlocal QNM analysis. For the major resonance peak, the responsible mode with nonlocal corrections is D5, which has a completely different electric field distribution from that of mode D of local response although their magnetic fields are essentially the same as shown in Fig.\,\ref{fig:fig2}(d). Moreover, instead of having a broad single resonance for the local case, the nonlocal spectral response is essentially comprised of the individual Lorentzian spectra of the involved QNMs, which have distinct electric field profiles. 

\emph{Conclusion.}---
We have formalized a theoretical framework of QNM analysis for general nonlocal media under linear mean-field approximation. The proposed GLM incorporates various kinds of specific nonlocal and quantum effects into a concise form and leads to a canonical definition of QNMs and to the construction of an orthonormalization scheme. The exemplary embodiments for QHDM and nonlocal polar dielectrics are explicitly shown. We applied the QHDM-based QNM analysis to reveal that the intrinsic bonding-to-tunneling mode transition in the quantum tunneling regime occurs at a smaller gap than inferred from far-field spectroscopic studies. Our work greatly facilitates analytical formulations of the electromagnetic responses in general nonlocal media and expands the application scenarios of QNM analysis for \textit{e.g.} Raman spectroscopy \cite{Zhang2013ChemicalMapping}, photon emissions from tunneling devices \cite{Novotny2019,SM} and molecular junctions \cite{lv2020}, and single photon emission from nanocavities \cite{Khurgin2020}. Therefore we hope it constitutes a valuable asset for nano-optics.

\begin{acknowledgments}
	We acknowledge financial support from the National Natural Science Foundation of China (Grant Number 9215011 and 11874166). We thank Wei Yan for useful discussions.
\end{acknowledgments}


\end{document}